\documentclass[aps,prd,groupedaddress,showpacs,showkeys]{revtex4}
\usepackage{latexsym,epsfig,amsmath,amssymb}
\usepackage{amssymb}
\usepackage{amsmath}
\usepackage{amsfonts}
\usepackage{epsfig} 
\usepackage{verbatim} 

\newcommand{\seq}{\begin{subequations}}
\newcommand{\sen}{\end{subequations}}
\newcommand{\eq}{\begin{eqnarray}}
\newcommand{\en}{\end{eqnarray}}
\newcommand{\ra}{\rangle}

\def\lc{\Lambda_c(2940)^+}
\def\lcp{\Lambda_c(2286)^+}
\def\lca{\Lambda_c}
\def\lcpa{\Lambda_c'}

\def\L2{\Lambda^2}

\begin{document}
\hfill KEK-TH-1359
\title{Radiative decay of $\Lambda_c(2940)^+$ 
in a hadronic molecule picture\\}  
\author{Yubing Dong$^{1,2}$, 
        Amand  Faessler$^3$,   
        Thomas Gutsche$^3$, 
        S. Kumano$^{4}$,
        Valery E. Lyubovitskij$^3$\footnote{On leave of absence
        from Department of Physics, Tomsk State University,
        634050 Tomsk, Russia}
\vspace*{1.2\baselineskip}}
\affiliation{
$^1$ Institute of High Energy Physics, Beijing 100049, P. R. China\\ 
\vspace*{.4\baselineskip} \\
$^2$ Theoretical Physics Center for Science Facilities (TPCSF), CAS, 
Beijing 100049, P. R. China\\ 
\vspace*{.4\baselineskip} \\ 
$^3$ Institut f\"ur Theoretische Physik,  Universit\"at T\"ubingen,\\
Kepler Center for Astro and Particle Physics, \\ 
Auf der Morgenstelle 14, D--72076 T\"ubingen, Germany\\
\vspace*{.4\baselineskip} \\ 
$^4$ 
KEK Theory Center, \\
Institute of Particle and Nuclear Studies, \\ 
High Energy Accelerator Research Organization (KEK), \\
and Department of Particle and Nuclear Studies, \\
Graduate University for Advanced Studies, \\
1-1, Ooho, Tsukuba, Ibaraki, 305-0801, Japan\\}

\date{\today}

\begin{abstract} 

The $\Lambda_c(2940)^+$ baryon with quantum numbers 
$J^P = \frac{1}{2}^+$ is considered as a molecular 
state composed of a nucleon and $D^\ast$ meson.
We give predictions for the width of the radiative decay process 
$\Lambda_c(2940)^+ \to \Lambda_c(2286)^+ + \gamma$ in this interpretation.  
Based on our results we argue that an experimental determination of the 
radiative decay width of $\lc$ is important
for the understanding of its intrinsic properties. 

\end{abstract}



\maketitle

\section{Introduction}

The new meson states of the $X$, $Y$ and $Z$ families,
which are strongly coupled to $c\bar c$ quark pairs, were dominantly detected
in B meson decays.
At the same time, in the analysis of $\Upsilon(4S)$ decay channels
two new charmed baryons ($C=+1$) denoted $\lc$ and $\Sigma_c(2800)$ were
discovered. The first one of these resonances was observed by 
the {\it BABAR} Collaboration~\cite{Aubert:2006sp} and later confirmed
by Belle~\cite{Abe:2006rz} as a resonant structure in the final state 
$\Sigma_c(2455)^{0,++} \pi^\pm \to \Lambda_c^+ \pi^+ \pi^-$ 
based on a 553 fb$^{-1}$ data sample 
collected at or near the $\Upsilon(4S)$ resonance at the KEKB collider. 
Both collaborations deduce values for mass and width with
$m_{\Lambda_c} = 2939.8 \pm 1.3 \pm 1.0$ MeV,
$\Gamma_{\Lambda_c} = 17.5 \pm 5.2 \pm 5.9$~MeV 
({\it BABAR}~\cite{Aubert:2006sp}) and 
$m_{\Lambda_c} = 2938.0 \pm 1.3^{+2.0}_{-4.0}$ MeV,
$\Gamma_{\Lambda_c} = 13^{+8 \ + 27}_{-5 \ -7}$ MeV 
(Belle~\cite{Abe:2006rz}) which are consistent with each other.

Concerning the $\lc$ some theoretical interpretations for this new charmed
baryon resonance were already discussed in the literature. For example,
in Ref.~\cite{He:2006is} the $\lc$ was regarded as a $D^{\ast 0} p$ molecular
state with its spin--parity being $J^P = \frac{1}{2}^-$ or $\frac{3}{2}^-$. 
This is due to the fact that the $\lc$ mass is just a few MeV below the
$D^{\ast 0} p$ threshold value.
It was shown that the boson-exchange mechanism, involving the $\pi$, $\omega$ 
and $\rho$ mesons, can provide binding in such $D^{\ast 0} p$ 
configurations. But in a first variant of a unitary meson-baryon coupled
channel model~\cite{GarciaRecio:2008dp} the $\lc$ cannot be identified with
a dynamically generated resonance.
In a relativized quark model~\cite{Capstick:1986bm} a charmed 
baryon state with  $J^P = \frac{3}{2}^+$ or $\frac{5}{2}^-$ 
is predicted in the 2940 MeV mass region.
Based on a calculation of the strong decay modes in
the $^3P_0$ model~\cite{Chen:2007xf} the possibility for $\lc$ being
the first radial excitation of the $\lcp$ is excluded since the decay
$\lc \to D^0 p$ vanishes for this configuration.
However, the possibility of being a $D$--wave charmed 
baryon with $J^P = \frac{1}{2}^+$ or $\frac{3}{2}^+$ was shown 
to be favored. Related studies concerning a conventional three-quark
interpretation of the $\lc$ baryon can also be found in
Refs.~\cite{Ebert:2007nw,Zhong:2007gp,Cheng:2006dk,%
Gerasyuta:2007un,Roberts:2007ni,Valcarce:2008dr,%
Wang:2008vj,Chen:2009tm}. 

We also recently considered the $\lc$ as a possible molecular state
composed of a nucleon and a $D^\ast$ meson as based 
on the so-called compositeness condition~\cite{Dong:2009tg}.
Its strong partial decay widths for the decay channel $pD^0$ as well as
$\Sigma_c^{++}\pi^-$ and $\Sigma_c^0\pi^+$ were estimated applying the
two different spin-parity assignments  $J^P = \frac{1}{2}^+$ 
and $\frac{1}{2}^-$.
For $J^P = \frac{1}{2}^+$ the sum of partial widths is consistent with
present observation, while for $\frac{1}{2}^-$ a severe overestimate for the
total decay width is obtained.
Hence the choice of spin-parity $J^P = \frac{1}{2}^+$ is preferred in the
molecular interpretation.

The technique for describing and treating composite hadron systems 
has been developed in Refs.~\cite{Faessler:2007gv,Dong:2008gb,Dong:2009tg}, 
where the recently observed unusual hadron states 
(like $D_{s0}^\ast(2317)$, $D_{s1}(2460)$, $X(3872)$, $Y(3940)$, 
$Y(4140)$, $Z(4430)$, $\lc$,~$\Sigma_c(2800)$) are analyzed 
as hadronic molecules. 
The composite structure of these possible molecular states is set up 
by the compositeness condition $Z=0$~\cite{Weinberg:1962hj,%
Efimov:1993ei,Anikin:1995cf,Dong:2008mt} 
(see also Refs.~\cite{Faessler:2007gv,Dong:2008gb,Dong:2009tg}).  
This condition implies that the renormalization constant of the hadron 
wave function is set equal to zero or that the hadron exists as a bound 
state of its constituents. The compositeness condition was originally 
applied to the study of the deuteron as a bound state of proton and
neutron~\cite{Weinberg:1962hj,Dong:2008mt}. Then it was extensively used
in low--energy hadron phenomenology as the master equation for the
treatment of mesons and baryons as bound states of light and heavy
constituent quarks (see e.g. Refs.~\cite{Efimov:1993ei,Anikin:1995cf}). 
By constructing a phenomenological Lagrangian including the 
couplings of the bound state to its constituents and the constituents 
to other final state particles we evaluated meson--loop 
diagrams which describe the different decay modes of the molecular states 
(see details in~\cite{Faessler:2007gv,Dong:2008gb}).  

Here we continue our study of the $\lc$ properties considering its 
radiative decay $\lc \to \lcp + \gamma$ in the hadronic molecule approach
developed in our recent paper~\cite{Dong:2009tg}. In particular,
electromagnetic transitions are often useful for probing the internal
structure of hadrons~\cite{Faessler:2007gv,Dong:2008gb,Anikin:1995cf,sk}.
Based on this previous study we choose the prefered $J^P = \frac{1}{2}^+$
assignment. As for the radiative decays of single charmed baryons in general
in future one can also expect a measurement on the possible radiative decay 
of the $\lc $ baryon. Upcoming experimental facilities like 
a Super B factory at KEK or LHCb might provide first data in this 
direction. Presently data are available on radiative decays of similar 
hadronic compounds in the meson sector like $D_{s0}(2317)$ and $D_{s1}(2460)$
which are supposed to be molecular states composed of a heavy and 
a light meson --- $DK$ and $D^\ast K$ bound state. 

In the present paper we proceed as 
follows. In Sec.~II we briefly discuss the basic notions of our approach. 
We discuss the effective Lagrangian for the treatment of the $\lc$ baryon 
as a superposition of the $p D^{\ast 0}$ and $n D^{\ast +}$ molecular 
components. Moreover, we consider the radiative decay 
$\lc \to \lcp + \gamma$ in this section. In Sec.~III we present our 
numerical results and, finally, in Sec.~IV a short summary.  

\section{Approach} 

In this section we briefly discuss the formalism for the study of the
$\lc$ baryon. Here we adopt spin and parity 
quantum numbers for the $\lc$ with $J^P = \frac{1}{2}^{+}$, where
consistency with the observed strong decay width of the 
$\lc$ was achieved in a hadronic molecule interpretation~\cite{Dong:2009tg}. 
Following Ref.~\cite{He:2006is} we consider this state as a superposition 
of the molecular $p D^{\ast 0}$ and $n D^{\ast +}$ components with 
the adjustable mixing angle $\theta$: 
\eq\label{Xstate}
|\lc\ra =   \cos\theta \ | p D^{\ast 0} \ra \ 
        + \ \sin\theta \ | n D^{\ast +} \ra 
\, . 
\en 
The values $\sin\theta = 1/\sqrt{2}$, $\sin\theta = 0$ or  
$\sin\theta = 1$ correspond to the cases of ideal mixing, of 
a vanishing $n D^{\ast +}$ or $p D^{\ast 0}$ component, respectively. 
Since the observed mass value of the $\lc$ with
$m_{D^{\ast 0}} + m_p - m_{\lc} =  5.94$ MeV and
$m_{D^{\ast +}} + m_n - m_{\lc} = 10.54$ MeV lies closer to the
$p D^{\ast 0}$ than to the $n D^{\ast +}$ threshold, we might expect
that the $|p D^{\ast 0}\ra$ configuration is the leading component.
Therefore, the mixing angle $\theta$ should be relatively small 
and we will vary its value from 0$^0$ to 25$^0$.

Our approach is based on an effective interaction Lagrangian describing 
the coupling of the $\lc$ to its constituents. We propose a setup for 
the $\lc$ in analogy to mesons consisting of a heavy quark and light 
antiquark, i.e. the heavy $D^\ast$ meson defines the center of mass 
of the $\lc$ while the light nucleon surrounds the $D^\ast$. 
The distribution of the nucleon relative to the $D^\ast$ meson 
is described by the correlation function $\Phi(y^2)$ depending on 
the Jacobi coordinate $y$. The simplest form of such a Lagrangian reads 
\eq\label{Lagr_Lc}
{\cal L}_{\Lambda_c}(x) &=& 
\bar\Lambda_c^+(x) \, \gamma^\mu \int d^4y \, \Phi(y^2) \, 
\Big( g_{\Lambda_c}^0 \, 
\cos\theta \,D^{\ast 0}_\mu(x) \, p(x+y) 
    + g_{\Lambda_c}^+ \, 
\sin\theta \,D^{\ast +}_\mu(x) \, n(x+y) \Big) 
\ + \ {\rm H.c.} 
\,, 
\en 
where $g_{\Lambda_c}^+$ and $g_{\Lambda_c}^0$ are the 
coupling constants of $\lc$ to the molecular $nD^{\ast +}$ and $pD^{\ast 0}$ 
components. Here we explicitly include isospin breaking 
effects by taking into account the neutron-proton and the $D^+ - D^0$ mass
differences. 
Note that in our previous analysis~\cite{Dong:2009tg} of the strong $\lc$ 
decays we restricted to the isospin symmetric limit. 
A basic requirement for the choice of an explicit form of the correlation 
function $\Phi(y^2)$ is that its Fourier transform vanishes sufficiently 
fast in the ultraviolet region of Euclidean space to render the Feynman 
diagrams ultraviolet finite. We adopt a Gaussian form for the correlation 
function. The Fourier transform of this vertex is given by
\eq 
\tilde\Phi(p_E^2/\Lambda^2) \doteq \exp( - p_E^2/\Lambda^2)\,,
\en 
where $p_{E}$ is the Euclidean Jacobi momentum. Here, 
$\Lambda$ is a size parameter characterizing 
the distribution of the nucleon in the $\lc$ baryon, which is a 
free model parameter regularizing the ultraviolet divergences of the 
Feynman diagrams. From the analysis of the strong decays of the 
$\lc$ baryon we found that $\Lambda \sim 1$ GeV~\cite{Dong:2009tg}. 
We might also expect that the $\lc$ is a quite compact molecular state
which, for example, is bound by exchange of a relatively massive hadron 
e.g. the scalar $f_0(600)$. Note that similar scale parameters were also 
obtained in the analysis of strong and radiative decay data of possible 
heavy-light hadronic molecules $D_{s0}(2317)=(DK)$ and 
$D_{s1}(2460)=(D^\ast K)$~\cite{Faessler:2007gv}. 
In the present analysis of the radiative decay of the $\lc$ we vary the 
size parameter $\Lambda$ in a wide range around this central value. 

In the kinematics we first restrict to the heavy quark limit (HQL)  
$m_{D^\ast} \to \infty$ supposing that the $D^\ast$ meson is 
located in the c.m. of the $\lc$. It is known that in the charm sector
the HQL is not always a good approximation due to possible, sizable
power corrections (in our case $m_N/m_{D^\ast}$). 
In the numerical analysis we will estimate how large these corrections are. 

The coupling constants $g_{\Lambda_c}^+$ and $g_{\Lambda_c}^0$ are the 
are determined by the 
compositeness condition~\cite{Weinberg:1962hj,Efimov:1993ei,Anikin:1995cf,%
Dong:2009tg,Faessler:2007gv}. It implies that the renormalization constant 
of the hadron wave function is set equal to zero with:
\eq\label{ZLc} 
Z_{\Lambda_c} = 1 - \Sigma_{\Lambda_c}^\prime(m_{\Lambda_c}) = 0 \,.
\en
Here, $\Sigma_{\Lambda_c}^\prime(m_{\Lambda_c})$   
is the derivative of the $\lc$ mass operator shown in Fig.1 and 
given by the expression: 
\eq\label{sigma_lc} 
\Sigma_{\Lambda_c}(p) = 
  (g_{\Lambda_c}^0)^2 \cos^2\theta \ \Pi_{pD^{\ast 0}}(p) 
+ (g_{\Lambda_c}^+)^2 \sin^2\theta \ \Pi_{nD^{\ast +}}(p) 
\en 
where $\Pi_{pD^{\ast 0}}(p)$ and $\Pi_{nD^{\ast +}}(p)$ are 
the loop integrals corresponding to the $pD^{\ast 0}$ and 
$nD^{\ast +}$ components, respectively. Therefore, the coupling 
constant $g_{\Lambda_c}^0$ is fixed from Eq.~(\ref{ZLc}) 
at the limit $\theta = 0^0$, 
while the $g_{\Lambda_c}^+$ is fixed from the same equation 
at the limit $\theta = 90^0$. 
Feynman diagrams contributing to the radiative decay of the $\lc$ in 
the hadronic molecule approach are shown in Fig.2. The $\lc \gamma $ final
state is fed by hadron loops containing the $\lc $ constituents.
Fig.2(a) stands for the direct coupling 
of the photon to the nucleon. The diagrams of Figs.2(b) and 2(c) are generated 
by the coupling of the photon to $D^\ast D$ and $D^\ast D^\ast$ 
meson pairs, respectively. The graph of Fig.2(d) is generated by
gauging the nonlocal strong interaction Lagrangian of Eq.~(\ref{Lagr_Lc}).
Finally, the pole diagrams in Figs.2(e) and 2(f) originate in the
direct coupling of the photon to $\lc$ and $\lcp$. Note, for a 
real photon the pole diagrams vanish due to gauge invariance. 

The phenomenological Lagrangian responsible for the full set of diagrams
in Fig.2 contains the coupling of $\lc$ to its constituents
(as already expressed in Eq.~(\ref{Lagr_Lc})) and the strong or
electromagnetic interaction Lagrangians involving these constituents coupled
to other fields in the loop or in the final state. These relevant
interaction vertices will be defined and discussed in the following. 
The electromagnetic part of the Lagrangian includes the following terms: 

\noindent 1) $NN\gamma$ interaction
which includes both minimal and nonminimal couplings  
\begin{eqnarray}
{\cal L}_{NN\gamma}(x)=e\bar{N}(x) \, 
\Big [ \, A_\mu(x) \, \gamma^\mu Q_{N} 
+ F_{\mu\nu}(x) \, \sigma^{\mu\nu} \, \frac{k_N}{4M_N} \, 
\Big ] \, N(x) \, , 
\end{eqnarray} 
2) $\lca\lca\gamma$ and $\lcpa\lcpa\gamma$ interaction Lagrangian  
(here and in the following by $\lca$ and $\lcpa$ 
denote the parent and daughter charmed baryons $\lc$ and $\lcp$):    
\begin{eqnarray}
{\cal L}_{\Lambda \Lambda \gamma}(x)
=e \sum\limits_{\Lambda = \Lambda_c, \Lambda_c'} \, 
\bar\Lambda(x) \, A_\mu(x) \, \gamma^\mu \, \Lambda(x) \, ,  
\end{eqnarray} 
3) $D^\ast D^\ast\gamma$ interaction is derived via minimal substitution 
in the free Lagrangian for charged $D^{\ast\pm}$ mesons 
\begin{eqnarray}
{\cal L}_{D^\ast D^\ast\gamma}(x)=i e A_\mu(x) \Big(
  g^{\mu\beta} D^{\ast -}_\alpha(x) \partial^\alpha D^{\ast +}_\beta(x) 
- g^{\alpha\beta} D^{\ast -}_\alpha(x) \partial^\mu D^{\ast +}_\beta(x) \Big) 
+ {\rm H.c.} \,,  
\end{eqnarray}
4) $D^\ast D\gamma$ 
interaction, which contains the nonminimal coupling 
$g_{D^\ast D\gamma}$ defining 
the decay rate $\Gamma(D^\ast \to D\gamma)$ (see e.g. discussion in 
Ref.~\cite{Dong:2008gb})
\begin{eqnarray}
{\cal L}_{D^\ast D\gamma}(x)=\frac{e}{4}g_{D^\ast D\gamma}
\epsilon^{\mu\nu\alpha\beta}F_{\mu\nu}(x)
\bar{D}^\ast_{\alpha\beta}(x)D(x) + {\rm H.c.} \,, 
\end{eqnarray} 
5) $\Lambda_c p D^\ast \gamma$ interaction Lagrangian
\eq\label{Lagr_Lc_em}
{\cal L}_{\Lambda_c p D^\ast \gamma} 
(x) &=& i e \, g_{\Lambda_c}^0 \, 
\cos\theta \, \bar\Lambda_c^+(x) \, \gamma^\mu \, D^{\ast 0}_\mu(x) 
\, \int d^4y \, \Phi(y^2) 
\, \int\limits^x_{x+y} dz_{\nu} \, A^{\nu}(z) \, p(x+y) 
\ + \ {\rm H.c.} \,, 
\en 
which is generated when gauging the nonlocal 
Lagrangian ${\cal L}_{\Lambda_c}$. In particular, to restore 
electromagnetic gauge invariance in ${\cal L}_{\Lambda_c}$, the proton 
field should be multiplied by the gauge 
field exponential (see further details 
in Refs.~\cite{Anikin:1995cf,Dong:2008mt}): 
\begin{eqnarray}
p(x+y)\to e^{ieI(x,x+y,P)}p(x+y) 
\,,~~~~~I(x,x+y,P)=\int\limits^x_{x+y} dz_{\nu}A^{\nu}(z).
\end{eqnarray}
For the derivative of $I(x,x+y,P)$ we use the path-independent prescription 
suggested in Ref.~\cite{Mandelstam:1962mi} which in turn states that the 
derivative of $I(x,x+y,P)$ does not depend on the path P originally used in 
the definition. The non-minimal substitution is therefore completely 
equivalent to the minimal prescription. Expanding the exponential term
of $e^{ieI(x,x+y,P)}$ in powers of the electromagnetic field and keeping the 
linear one, we derive the Lagrangian (\ref{Lagr_Lc_em}) and therefore 
generate the vertex contained in the diagram of Fig.2(d).

In the preceding expressions we introduced several notations.  
$Q_N$ and $k_N$ are the nucleon charge and anomalous magnetic moments: 
$Q_p = 1$, $Q_n = 0$, $k_p = 1.793$, $k_n = -1.913$. 
$A_{\mu}$ is the photon field. 
$F_{\mu\nu} = \partial_\mu A_\nu - \partial_\nu A_\mu$ 
and $D^\ast_{\alpha\beta} =\partial_{\alpha}D^{\ast}_{\beta}
-\partial_{\beta}D^{\ast}_{\alpha}$ 
are the stress tensors of the electromagnetic field and 
$D^\ast$, respectively.  
The coupling constant $g_{D^\ast D\gamma}$ is fixed by data (central
values) on the radiative decay widths
$\Gamma(D^\ast \to D\gamma)$~\cite{PDG:2008}: 
\eq
g_{D^{\ast \pm}D^\pm\gamma} = 0.5 \ {\rm GeV}^{-1}\,, \hspace*{.5cm}
g_{D^{\ast 0}D^0\gamma} = 2.0 \ {\rm GeV}^{-1}\,. \hspace*{.5cm}
\en

The relevant strong interaction Lagrangian contains two types of couplings --- 
$N D^\ast \lcpa$ and $N D \lcpa$:  
\begin{eqnarray}
{\cal L}_{ND^\ast\Lambda_c'}=g_{N D^\ast \Lambda_c'}
\bar{N}\gamma^{\mu}\Lambda'_c \bar D^\ast_{\mu} \ + \ {\rm H.c.} 
\end{eqnarray}
and 
\begin{eqnarray}
{\cal L}_{ND\Lambda_c'}=g_{ND\Lambda'_{c}}
\bar{N}i \gamma_5\Lambda'_{c} \bar D \ + \ {\rm H.c.} 
\end{eqnarray}
The couplings $g_{N D^\ast \Lambda'_c}$ and $g_{ND\Lambda'_{c}}$ 
can be deduced from the phenomenological flavor-SU(4)
Lagrangian~\cite{Dong:2009tg,Okubo:1975sc} with 
\begin{eqnarray}
g_{ND\Lambda'_{c}}=-\frac{3\sqrt{3}}{5}g_{\pi NN}\; , \;
g_{ND^\ast\Lambda'_{c}}= - \frac{\sqrt{3}}{2}g_{\rho NN} \; ,
\end{eqnarray}
expressed in terms of the $\pi NN$ and $\rho NN $ couplings with values
\begin{eqnarray}
g_{\rho NN}=6 \;, \; g_{\pi NN}=13.2 \; .
\end{eqnarray}

For the calculation of the electromagnetic transition amplitude between 
the two spin-$\frac{1}{2}$ particles $\Lambda_c$ and $\Lambda_c'$ we have 
to consider the constraints of gauge invariance. 
In case of a general off-shell one-photon transition the invariant
matrix element reads as
\eq
{\cal M}^\mu(p,p') = 
\bar u_{\Lambda_c'}(p')
\Gamma^\mu(p,p') 
u_{\Lambda_c}(p)\,,   
\en 
where the vertex function $\Gamma^{\mu}(p,p')$ is decomposed in terms of 
three relativistic form factors $F_{1,2,3}(q^2)$ with the structure
\eq 
\Gamma^{\mu}(p,p')
=F_1(q^2)\gamma^{\mu}+F_{2}(q^2)i\sigma^{\mu\nu}q_{\nu}+F_3(q^2)q^{\mu} \,.
\en 
Here $u_{\lcpa}(p')$ and $u_{\lca}(p)$ are 
the spinors of daughter and parent baryons, respectively. 
Due to gauge invariance with $q_\mu {\cal M}^{\mu}(p,p') = 0$
the form factors $F_1(q^2)$ and $F_3(q^2)$ are related as 
\eq 
F_1(q^2)=F_3(q^2)\frac{q^2}{m_{\Lambda_c}-m_{\Lambda_c'}}.
\en 
Therefore, in the limiting case of a real photon ($q^2 = 0$) the 
invariant matrix element of the $\lc \to \lcp + \gamma$ transition 
is expressed in terms of the spin-flip form factor only with 
\eq 
{\cal M}^\mu(p,p')= \frac{F_{\lca\lcpa\gamma}}{2m_{\lca}}  \, 
\bar u_{\lcpa}(p') \, i \, \sigma^{\mu\nu} \, q_\nu \, u_{\lca}(p) \, . 
\en 
The coefficient $F_{\lca\lcpa\gamma} \equiv 2m_{\lca}  F_2(0)$ 
is the effective
coupling of $\lc\lcp\gamma$, deduced from the set of graphs of Fig.~2,
determined in our approach. 
This effective coupling contains the loop integrals which are
evaluated using the calculational techniques developed and 
explicitly shown in Refs.~\cite{Dong:2009tg}-\cite{Dong:2008gb}.  
Once this effective coupling is determined the final expression for
the decay width is given by 
\eq 
\Gamma(\lc \to \lcp + \gamma) = 
\frac{\alpha P^{\ast 3}}{m_{\Lambda_c}^2} \, F_{\lca\lcpa\gamma}^2 \; ,
\en
where $P^\ast = (m_{\lca}^2-m_{\lcpa}^2)/(2m_{\lca})$ is 
the three-momentum of the decay products in the rest frame of 
the initial $\lc$ baryon.

\section{Numerical results}

For our numerical calculations the input masses of $D^{\ast 0}$, 
$D^{\ast +}$, $p$, $n$, $\lc$ and $\lcp$ are taken from the compilation
of the Particle Data Group~\cite{PDG:2008}. The only free parameter in
our calculation is the dimensional parameter~$\Lambda$. As already stated,
this parameter describes the distribution of the nucleon around the
$D^\ast$ which is located in the center-of-mass of the $\lc$.
Here we select $\Lambda\sim 1$~GeV, a value which is close to the scale
set by the nucleon mass as usually taken in hadronic
interactions~\cite{sk-form}.
In the calculation we consider a variation of this value
from 0.25 to 1.25 GeV. 

In Table I we first show the dependence of the calculated couplings 
$g_{\Lambda_c}^0$ and $g_{\Lambda_c}^+$ 
on this free parameter $\Lambda$, which are fixed using the 
compositeness condition [see Eqs.~(\ref{ZLc}) and (\ref{sigma_lc})]. 
We find that the difference in the binding 
energies ($M_p+M_{D^{\ast 0}}-m_{\lc} = 5.9$~MeV and 
$M_n+M_{D^{\ast +}}-m_{\lc} = 10.5$~MeV) leads to some deviation between
the respective coupling constants $g_{\Lambda_c}^0$ and $g_{\Lambda_c}^+$. 
Also decreasing of the scale parameter $\Lambda$ leads to decreasing 
of the couplings $g_{\Lambda_c}^0$ and $g_{\Lambda_c}^+$. One can see 
that at values of $\Lambda \leq 0.75$ GeV the couplings are quite 
suppressed, so the preferred region for the fixing parameters $\Lambda$ is 
around 1 GeV.  In Tables II and III we present the numerical results 
for the effective coupling 
constant $F_{\lca\lcpa\gamma}$ and for the resulting radiative decay width
of the process $\lc \rightarrow \lcp + \gamma$. The predictions for the
decay width are given for selected values of 
$\Lambda = 0.25$, $0.4$, $0.5$, $0.75$, $1$, $1.25$~GeV 
and for a variety of mixing angles $\theta$ in the interval $(0 - 25)^0$.  
Our results are rather sensitive to a variation of the scale parameter 
$\Lambda$. This should be obvious since the ultraviolet divergence 
of the diagrams is regularized by the cutoff~$\Lambda$. Again at relatively 
small values of $\Lambda$ the predictions for the decay parameters are 
very small. 
The results also possess a pronounced sensitivity on a variation of 
the mixing parameter $\theta$. An increase of $\theta$ leads to a suppression
of the effective coupling and hence the decay width. The range of the estimated
decay width is rather wide and varies from several to hundred keV. 
This is mainly due to the nontrivial cancellation between diagrams 
involving the $|p D^{\ast 0}\ra$ and the $|n D^{\ast +}\ra$ components 
in the loops. For illustration of this behavior in Table~4 we present 
the contributions of the different 
diagrams to the effective coupling at values of $\Lambda = 1$ GeV and 
$\theta = 10^0$. 
All contributions involving the $|n D^{\ast +}\ra$ 
component are destructive in comparison to  the leading contribution 
giving by the $|p D^{\ast 0}\ra$ component in the diagram of Fig.2(a). 
This leads to a suppression of the effective coupling and the width when 
the fraction of the $|n D^{\ast +}\ra$ component (or the value of 
the mixing angle $\theta$) is increased. Our final comment concerns 
an estimate of power corrections to the decay rate due to the shift 
of the position of the $D^\ast$ from the $\lc$ c.m. These corrections 
depend on the scale parameter $\Lambda$. We found that for
$\Lambda = 1$ GeV these corrections are up to 10\% depending on 
the mixing angle $\theta$. Varying $\Lambda$ 
from 1 GeV to 0.25 GeV these corrections increase up to 30\%, while
they are reduced when $\Lambda$ increases. For completeness we present 
the results including power corrections for the specific values of the 
model parameters $\Lambda = 1$ GeV and $\theta = 10^0$. They are given 
in Table~4 in brackets. 

\section{Summary}

To summarize, we pursue a hadronic molecule interpretation of the recently 
observed charmed baryon $\lc$ studying its consequences for the radiative 
decay mode $\lcp  \gamma$ for spin-parity $J^P=\frac{1}{2}^+$.
In the present scenario the 
$\lc$ baryon is described by a superposition of 
$|p D^{\ast 0}\ra$ and $|n D^{\ast +}\ra$ components with the explicit 
admixture expressed by the mixing angle $\theta$.
Our numerical results for the radiative decay widths show 
that the contribution of diagram Fig.2(a) gives the 
leading contribution while those of Figs. 2(b), 2(c) and 2(d) are subleading
but non-negligible.
The diagrams of Fig.2(e) and 2(f) vanish for real photons and, therefore, 
do not contribute to the process $\Gamma(\lc \to \lcp + \gamma)$. 
The calculated radiative decay widths display a sizable sensitivity to
the mixing angle $\theta$ and to the scale parameter $\Lambda$.  
Especially the cancellation between the contributions of the 
diagrams Figs.2(a)-2(d) results in a rather pronounced $\theta$-dependence. 
This effect can provide a stringent constraint on the role of the two  
molecular components $pD^{\ast 0}$ and $nD^{\ast +}$ in the $\lc$ resonance. 
Possible future measurements of the radiative decay  
width can provide further insights into the structure of the $\lc$ state.
New facilities like the Super B factory at KEK 
or LHCb might have the capability to reach
the sensitivity to detect radiative decays of charmed baryons in the keV
regime.

\begin{acknowledgments}

This work is supported  by the National Sciences Foundations 
No. 10775148, 10975146, by the CAS grant No. KJCX3-SYW-N2 (YBD) 
and by the DFG under Contract No. FA67/31-2 and No. GRK683. 
This research is also part of the European
Community-Research Infrastructure Integrating Activity
``Study of Strongly Interacting Matter'' (HadronPhysics2,
Grant Agreement No. 227431), Russian President grant
``Scientific Schools''  No. 3400.2010.2, Russian Science and
Innovations Federal Agency contract No. 02.740.11.0238. 
Author (YBD) thanks the theory group of KEK, Japan for the hospitality. 
The authors thank R. Mizuk for comments on experimental possibilities. 

\end{acknowledgments}

\newpage 

\begin{figure}
\centering{\
\epsfig{figure=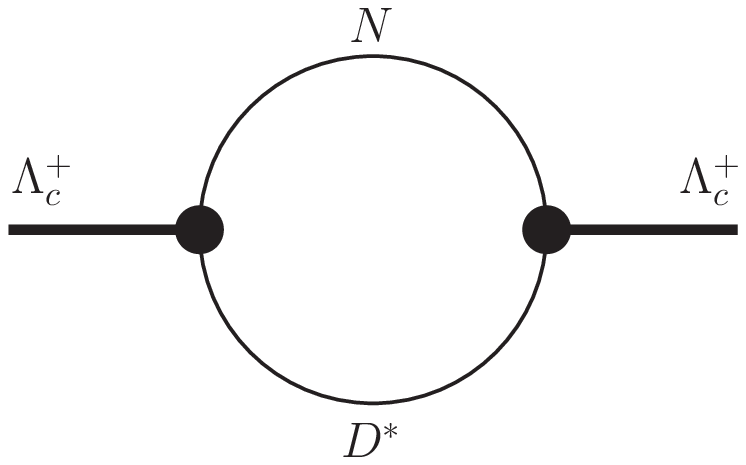,scale=.65}}
\caption{Diagram describing the $\lc$ mass operator.}
\label{fig:str}
%\end{figure}

\vspace*{.5cm}
%\begin{figure}
\centering{\
\epsfig{figure=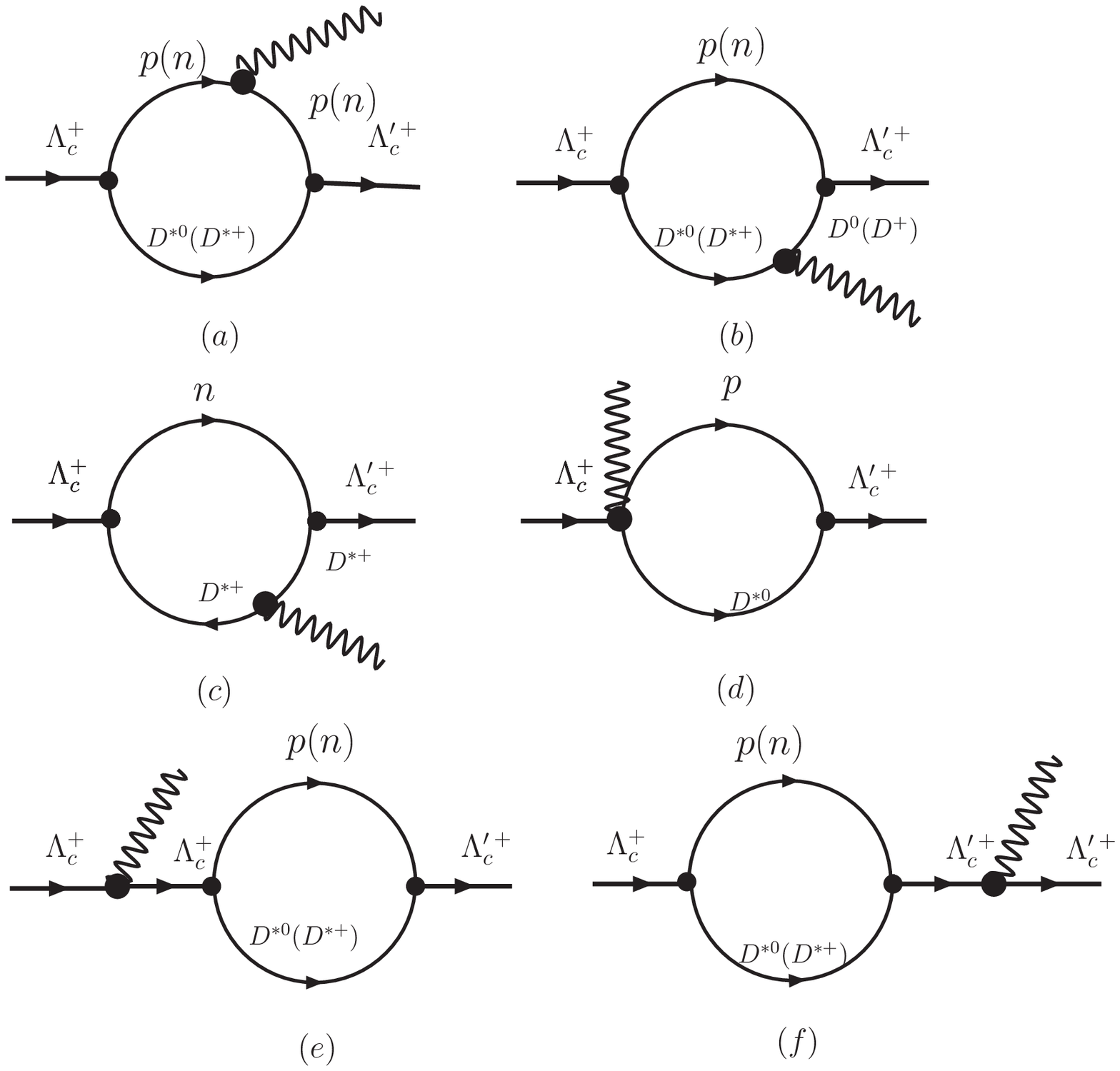,scale=.65}}
\caption{Diagrams contributing to the
radiative decay process $\lc \to \lcp \gamma$.}
\label{fig:vertex}
\end{figure}

\newpage

\begin{table}
\begin{center}
{\bf Table I.} Couplings $g_{\Lambda_c}^0$ and $g_{\Lambda_c}^+$.
 
\vspace*{.25cm}
\def\arraystretch{1.4}
\begin{tabular}{|c|c|c|}  
\hline 
    $\Lambda$ (GeV)
    & $g_{\Lambda_c}^0$ 
    & $g_{\Lambda_c}^+$ \\ 
\hline 
0.25 & 3 $\times 10^{-5}$   & 4 $\times 10^{-5}$    \\ \hline
0.4  & 1.2 $\times 10^{-2}$ & 1.6 $\times 10^{-2}$ \\ \hline
0.5  & 0.09 & 0.10 \\ \hline
0.75 & 0.56 & 0.67 \\ \hline
1    & 1.09 & 1.29 \\ \hline
1.25 & 1.51 & 1.74 \\ \hline 
\end{tabular}
\end{center}
%\end{table} 

\vspace*{.5cm}
%\begin{table}
\begin{center}
{\bf Table II.} Effective coupling constant $F_{\lca\lcpa\gamma}$. 

\vspace*{.25cm}
\def\arraystretch{1.4}
\begin{tabular}{|c|c|c|c|c|c|c|}  
\hline 
$\theta$ 
& \multicolumn{6}{c|}{$\Lambda$ (GeV)}\\
\cline{2-7}
(in grad)   & $0.25$ & $0.4$ & $0.5$ & $0.75$
            & $1$
            & $1.25$ \\
\hline 
0  & 0.26 & 0.46 & 0.61 & 0.83 & 0.93 & 0.97 \\ 
5  & 0.24 & 0.42 & 0.56 & 0.74 & 0.82 & 0.85 \\
10 & 0.22 & 0.37 & 0.50 & 0.66 & 0.71 & 0.72 \\ 
15 & 0.21 & 0.33 & 0.43 & 0.56 & 0.60 & 0.58 \\ 
20 & 0.16 & 0.28 & 0.37 & 0.46 & 0.47 & 0.44 \\ 
25 & 0.13 & 0.22 & 0.30 & 0.36 & 0.35 & 0.30 \\
\hline
\end{tabular}
\end{center}
%\end{table} 

\vspace*{.5cm}

%\begin{table}
\begin{center}
{\bf Table III.} 
Radiative decay width of $\lc$ in keV. 

\vspace*{.25cm}
\def\arraystretch{1.4}
\begin{tabular}{|c|c|c|c|c|c|c|}  
\hline 
$\theta$ & \multicolumn{6}{c|}{$\Lambda$ (GeV)}\\
\cline{2-7}
(in grad)   & $0.25$ & $0.4$ & $0.5$ & $0.75$
            & $1$
            & $1.25$ \\
\hline 
0  & 11.1 & 35.4 & 61.7 & 113.1 & 142.7 & 156.8 \\
5  &  9.2 & 29.2 & 51.0 & 91.5 & 112.2 & 119.4 \\ 
10 &  7.4 & 23.2 & 40.6 & 71.0 & 83.9  & 85.5  \\
15 &  5.7 & 17.6 & 30.8 & 52.1 & 58.6  & 56.2  \\
20 &  4.1 & 12.5 & 22.0 & 35.5 & 37.1  & 32.4  \\
25 &  2.7 & 8.1  & 14.4 & 21.7 & 20.1  & 14.7  \\
\hline
\end{tabular}
\end{center}
%\end{table} 

\vspace*{.5cm}

\newpage

%\begin{table}
\begin{center}
{\bf Table IV.} 
Contributions of the diagrams Figs. 2(a)-(d) to 
$F_{\lca\lcpa\gamma}$ at $\Lambda = 1$ GeV and $\theta = 10^0$. \\
Numbers in brackets include power corrections discussed in the text. 

\vspace*{.25cm}
\def\arraystretch{1.4}
\begin{tabular}{|c|r|r|r|}  
\hline 
Diagram & \multicolumn{3}{c|}{$F_{\lca\lcpa\gamma}$}\\
\cline{2-4}
    & $p D^{\ast 0}$ 
    & $n D^{\ast +}$
    & $p D^{\ast 0} + n D^{\ast +}$ \\
\hline 
Fig.2(a)  &     1.00 (1.17) & $-$ 0.16 ($-$0.40) &     0.84 (0.77) 
\hspace*{.6cm} \\
Fig.2(b)  & $-$ 0.13 ($-$ 0.25) & $-$ 0.01 ($-$ 0.01) 
& $-$ 0.14 ($-$ 0.26)   \hspace*{.6cm} \\
Fig.2(c)  &     0 (0)   & $-$ 0.04 ($-$ 0.04)  & $-$ 0.04 ($-$ 0.04) 
\hspace*{.6cm} \\
Fig.2(d)  &     0.05 (0.13) & 0 (0.08) & 0.05 (0.21) \hspace*{.6cm} \\
Total     &     0.92 (1.05) & $-$ 0.21 ($-$ 0.37)  
&     0.71 (0.68) \hspace*{.6cm} \\
\hline
\end{tabular}
\end{center}
\end{table} 

\end{document}